\documentclass[epj]{svjour}
\usepackage{graphics}

\newcommand{\btem}{\bibitem}
\newcommand{\beq}{\begin{eqnarray}}
\newcommand{\eeq}{\end{eqnarray}}

\renewcommand{\d}{\partial}
\newcommand{\eps}{\epsilon}

\newcommand{\bfq}{\mbox{{\boldmath $q$}}}

\newcommand{\bfF}{\mbox{{\boldmath $F$}}}

\def\Vec#1{\mbox{\boldmath$#1$}}
\def\Ren#1{\mbox{\boldmath$#1$}}

\begin{document}
\title{
Derivation of relativistic hydrodynamic equations consistent 
with relativistic Boltzmann equation by renormalization-group
method}
\author{Kyosuke Tsumura\inst{1} 
\and Teiji Kunihiro\inst{2}
}                     
%
%
\institute{Analysis Technology Center,
  Fujifilm Corporation,  Kanagawa 250-0193, Japan \and 
Department of Physics,
  Kyoto University,   Kyoto 606-8502, Japan}
\date{Received: date / Revised version: date}
%
\abstract{
We review our work on the application of the renormalization-group method to obtain 
first- and second-order relativistic hydrodynamics
of the relativistic Boltzmann equation (RBE) as a dynamical system,
with some corrections and new unpublished results.
For the first-order equation,
we explicitly obtain the distribution function in the 
asymptotic regime as the invariant manifold of the dynamical system,
which turns out to be nothing but the matching condition defining the energy frame, \textit{i.e.}, the Landau-Lifshitz one.
It is argued that the frame on which the flow of the relativistic hydrodynamic equation
 is defined must be the energy frame,
if the dynamics should be consistent with the underlying RBE. 
A sketch is also given for derivation of the second-order hydrodynamic equation,
{\it i.e.}, extended thermodynamics, which is accomplished by extending the 
invariant manifold so that it is spanned by excited modes as well as the
zero modes (hydrodynamic modes) of the linearized collision operator.
 On the basis of thus constructed resummed distribution function,
we propose a novel ansatz for the functional form to be used in 
Grad moment method; it is shown that our theory gives the same
expressions for the transport coefficients 
as those given in the Chapman-Enskog theory as well as the
novel expressions for the relaxation times and lengths allowing natural 
interpretation.
\PACS{
      {PACS-key}{describing text of that key}   \and
      {PACS-key}{describing text of that key}
     } 
} 
\maketitle
\renewcommand{\theequation}{\arabic{section}.\arabic{equation}}

\setcounter{equation}{0}
\section{
   Introduction
}
\label{intro}

The dynamical evolution of
the hot and/or dense QCD matter
produced in the Relativistic Heavy Ion Collider (RHIC) at Brookhaven National Laboratory
can be well described
by relativistic hydrodynamic simulations \cite{qcd001,qcd003}.
It seems to be the case also 
for the created matter in LHC (Large Hadron Collider) in European Organization 
for Nuclear Research (CERN); see, for example, \cite{Bozek:2011gq,Hirano:2012kj}. 
The suggestion that the created matter at RHIC may have only a tiny viscosity
prompted an interest in the origin of the viscosity in the
created matter to be described using the relativistic quantum field theory
and also the dissipative hydrodynamic equations.
We note that since the created matter expands,
the proper dynamics for the description may change from hydrodynamics to kinetic one and vice versa
\cite{Hirano:2012kj,connect-1,connect-2,Hirano:2005wx,dis005,nonaka06}.
 The hydrodynamics is also relevant to the soft-mode dynamics 
\cite{Fujii:2004jt,Son:2004iv,Minami:2009hn} around
the possible critical point(s) in QCD phase diagram \cite{asakawa89};
see \cite{Zhang:2011xi} for the latest up date.

  However,
  the theory of relativistic hydrodynamics for viscous fluids is still 
under debate.
  In fact, we can indicate the following problems:
  (1)~ There are ambiguities  in the definition of the flow velocity
  \cite{hen001,hen002,van-and-biro2012,osada2012}.
  (2)~ In the Eckart (particle) frame, 
  there arises an unphysical instabilities of the equilibrium state \cite{hyd002}.
  (3)~ The so called first-order equations lack in  causality, 
{\it i.e.}, some components of the hydrodynamic equations are of parabolic nature
\cite{mic001,mic001-2,hen003,mic004}.

  Taking
 the relativistic Boltzmann equation (RBE) \cite{mic001,mic001-2} as a typical kinetic equation,
 we have been exploring
 the basic problems with the relativistic hydrodynamics 
\cite{env009,Tsumura:2009vm,Tsumura:2011cj}.
We note that 
such an approach is important also for a systematic analysis of  RHIC/LHC data, because
  the proper dynamics for the description 
may change from hydrodynamics to kinetic one and vice versa, as mentioned above.

It is conjectured \cite{bogoliubov,text} that the non-equilibrium process evolves
  through  some stages of hierarchical  dynamics:
  In the beginning of the time evolution of an isolated prepared state,
  the whole dynamical evolution of the system
  will be governed by Hamiltonian dynamics
  that is time-reversal invariant.
  As the system gets old,
  the dynamics is relaxed into the kinetic regime
  where the time-evolution system is well described by
   kinetic equations which describe a coarse-grained slower dynamics:
 The Boltzmann equation for the one-body distribution function
  is one of them \cite{text}.
 Usually the original time-reversal invariance is lost in the description 
by the kinetic equation through the coarse-graining.
  As the system is further relaxed,
  the time evolution will be described in terms of the hydrodynamic quantities,
  \textit{i.e.},  the flow velocity,   particle-number density,
  and local temperature.
  In this sense,
  the hydrodynamics is the far-infrared asymptotic dynamics of
  the kinetic equation.

Thus, for obtaining the proper relativistic hydrodynamic equation, 
it is a legitimate and natural way to start with 
the RBE which is Lorentz invariant and expected to be free from causality problem 
\cite{mic001,mic001-2}; moreover, apparent instability is not known for numerical simulations
of the RBE, as far as we are aware of, 
and the stability is proved at least for the 
linearized version of it \cite{dudynski1985,strain2010}.
For analyzing the problems (1) and (2) first,
we derive hydrodynamic equation \cite{env009,Tsumura:2011cj} from the RBE.
We note  that the problem is a typical reduction problem of
a dynamical system in the far-infrared long-wave length limit.
So we need a powerful reduction theory for our purpose, and 
we shall take the renormalization-group (RG) method \cite{rgm001,env001,env006} as 
such a powerful one.
The reduction of dynamics can be viewed as a construction 
of an invariant/attractive manifold \cite{holmes,kuramoto}, 
and it has been shown \cite{env001,env006}
 that the RG method can be nicely formulated as an elementary
method for constructing the invariant manifold of a given dynamical system.

In this article, we also report on our attempt \cite{next002}
to examine the causality problem (3) by deriving the so called
 {\it extended thermodynamics} \cite{extended,meso,Gorban}: Namely, we derive  mesoscopic
dynamics of the RBE by constructing the invariant/attractive manifold incorporating
some fast modes
as well as the zero modes of the linearized collision operator.
It turns out that our theory leads to 
 the same expressions for the transport coefficients as given by the
Chapman-Enskog method \cite{chapman} and also novel formulas
 of the relaxation times in terms of relaxation functions,
which allow a natural physical interpretation of the relaxation times. 
Moreover, the distribution function which is explicitly constructed in our theory provides
a new ansatz for the functional form of the distribution function
in the Grad theory \cite{grad}.

 \setcounter{equation}{0}
  \section{
Introduction to renormalization-group method by an example
  }

 Our approach is heavily based on the reduction theory of dynamics called
the renormalization-group (RG) method \cite{rgm001,env001,env006}, and the
reliability of our theory is assured by that of the method.
It is nice \cite{env006} that the RG method can be formulated as an elementary 
way of construction of the  invariant/attractive  manifold of dynamical systems;
it not only 
leads to asymptotic dynamics of a given equation but also
extract explicitly the differential equations governing the would-be constants appearing in the
solution to the differential equation.
  In this section,  we  make an account of 
  the RG method 
  using a simple non-linear equation.

  Let us take the Van der Pol equation as an example:
  \begin{eqnarray}
    \ddot{x} + x = \eps \, (1 - x^2) \, \dot{x},
    \label{eq:van-der-pol}
  \end{eqnarray}
  with $\eps$ being small.

  Let $\tilde{x}(t;\,t_0)$ be a local solution
  around $t\sim \forall t_0$, and
  represent it as a perturbation series;
  \beq
\tilde{x}(t;\,t_0) = \tilde{x}_0(t;\,t_0)
  +\eps \, \tilde{x}_1(t;\,t_0) + \eps^2 \, \tilde{x}_2(t;\,t_0) + \cdots.
\eeq
  In the RG method,
  the initial value $W(t_0)$ is to  constitute the desired (approximate) solution in 
a global domain and make the invariant manifold of the system.
  We suppose that an exact solution is given by $x(t)$
  and the initial value of $\tilde{x}(t;\,t_0)$ at $t=t_0$ is set up to be $x(t_0)$,
  \textit{i.e.}, 
\beq
W(t_0)\equiv \tilde{x}(t_0;\,t_0) = x(t_0).
\eeq
  The initial value as the exact solution should also be expanded as
\beq
  W(t_0) = W_0(t_0) + \eps \, W_1(t_0) + \eps^2 \, W_2(t_0) + \cdots.
\eeq

  The zeroth-order equation reads
  \begin{eqnarray}
    {\cal L}\tilde{x}_0  = 0,
\end{eqnarray}
 with
${\cal L}_0 \equiv  \frac{d^2}{dt^2} + 1$.
The solution  may be expressed as
  \begin{eqnarray}
    \tilde{x}_0(t;\,t_0) = A(t_0) \, \cos(t + \theta(t_0)),
\end{eqnarray}
with the initial value $W(t_0)=\tilde{x}_0(t_0;\,t_0) = A(t_0) \, \cos(t_0 + \theta(t_0))$.
  Note that 
  the integral constants $A$ and $\theta$ may depend on the
  initial time $t_0$. The integration constants $A(t_0)$ and $\theta(t_0)$ will parametrize
the global solution and correspond to the hydrodynamic variables which parametrize
the distribution function in the local equilibrium.

  The equation for $\tilde{x}_1$ reads
  \begin{eqnarray}
    {\cal L}\tilde{x}_1 =
    - A \left( 1 - \frac{A^2}{4} \right) \sin\phi(t)
    + \frac{A^3}{4}  \sin 3\phi(t),
  \end{eqnarray}
  with $\phi(t) = t + \theta_0(t_0)$.
  Notice that the first term in r.h.s. is   a zero mode of  ${\cal L}$, and hence
 the special solution to this equation   contains a secular term
  that is expressed as a product of  $t$ and a zero mode of ${\cal L}$.
  Since we have supposed that   the initial value at $t=t_0$ is on an exact solution,
 we should make the corrections from the zeroth-order solution
  as small as possible.   This condition is realized by letting the secular terms
   vanish at $t=t_0$,
  which is possible because we can freely add zero mode solutions to
  a special solution.
  Thus,
  we have the first-order solution  as
  \begin{eqnarray}
    \tilde{x}_1(t;t_0) &=& (t-t_0) \frac{A}{2} \left(1 -
    \frac{A^2}{4}\right) \sin \phi(t)\nonumber\\
    &&{}- \frac{A^3}{32} \sin 3\phi(t),
  \end{eqnarray}
with the initial value at $t=t_0$,
$W_1(t_0)=\tilde{x}_1(t_0;t_0) = - {A^3(t_0)}/{32}\cdot \sin 3\phi(t_0)$.
 
  If we stop here, we have the perturbative solution;
 $\tilde{x} = \tilde{x}_0 + \eps \, \tilde{x}_1$,
  which should be valid in a local domain $t\sim t_0$ but 
becomes invalid in the global domain where 
 $\vert t-t_0 \vert$ can be large,
due to the secular term.

  We shall now take a geometrical  point of view \cite{env001}:
The function $\tilde{x}(t; t_0)$ corresponds to
  a curve drawn in the $(t,\,x)$ plane for each $t_0$;
  in other words,
  we have a {\em family of curves} represented by $\tilde{x}(t;t_0)$
  in the $(t,\,x)$ plane;
  a member of the family is parametrized by $t_0$, and
  each member is close to an exact solution in the neighborhood of
  $t = t_0$.
  Thus,
  an idea is that the {\em envelope curve} of the family of curves
  should give a global solution.
  The classical theory of envelope curve says that the envelope 
can be constructed by solving
  the following equation,
  \begin{eqnarray}
    \frac{d\tilde{x}}{dt_0} \Big \vert_{t_0=t} = 0,
    \label{eq:van-der-pol-env}
  \end{eqnarray}
  which is called the RG equation \cite{rgm001}; we have 
here made an account of it on the basis of the envelope theory \cite{env001}.
Eq. (\ref{eq:van-der-pol-env}) leads to the equations for $A(t)$ and $\phi(t)$,
  \begin{eqnarray}
    \dot{A} = \eps \frac{A}{2}  \left( 1-\frac{A^2}{4} \right),
 \quad    \dot{\phi} = 1,
\label{eq:amp-phase-eq}
  \end{eqnarray}
which equations may be called an amplitude and phase equation, respectively.
The original equation is reduced to these simpler equations for the amplitude and phase
which parametrize the solution of the original equation.
  These reduced equations are readily solved, with which a resummation of
the perturbation series is performed;
the resumed solution is found to successfully admit
 a limit cycle with a radius of $2$.

  The resultant envelope function as a global solution
  is given by 
  \begin{eqnarray}
    x_{\mathrm{E}}(t) &\equiv& \tilde{x}(t; t) = W(t) \nonumber \\
    &=& A(t)\,\cos\phi(t) - \eps \frac{A^3(t)}{32} 
    \sin 3\,\phi(t),
  \end{eqnarray}
  with $A(t)$ and $\phi(t)$ being the solution of Eq. (\ref{eq:amp-phase-eq}).
  Thus, 
  we have succeeded in not only obtaining
  the asymptotic solution as a whole but also extracting
  the slow variables $A(t)$ and $\phi(t)$
  explicitly and their governing equations.

 However,  there is a problem left:
Does $x_{\mathrm{E}}(t) \equiv \tilde{x}(t; t)$ indeed satisfy the original differential
equation (\ref{eq:van-der-pol})?
 We give here a proof for that \cite{env001}.

First let us rewrite Eq. (\ref{eq:van-der-pol}) into
a coupled equation of {\em first order}:
\beq
\frac{d{\bfq}(t)}{dt}={\bfF}({\bfq}(t); \eps),
\label{eq:coupled-van-der-pol}
\eeq
where 
\beq
\bfq &=&\, ^t(q_1=x, \, q_2=\dot{x}), \\
\bfF &=&\, ^t(q_2, \, -q_1 + \eps \, (1-q^2_1)\, q_2).
\eeq
 We  have an approximate local solution to Eq. (\ref{eq:coupled-van-der-pol}) 
 $\tilde{\bfq}(t; t_0)$ 
 around $t=t_0$ up to $O(\eps^n)$, corresponding to $\tilde{x}(t; t_0)$;
\beq
\frac{d\tilde{{\bfq}}}{dt}={\bfF}(\tilde{\bfq}(t; t_0); \eps) + O(\eps^n).
\label{eq:van-der-coupl-approx}
\eeq
Now, the RG/envelope equation implies that 
\beq
\frac{\partial \tilde{\bfq}(t; t_0)}{\partial t_0}\Big\vert_{t=t_0}=0.
\label{eq:rg-envelope-van-der}
\eeq
The envelope function  $\bfq_{\mathrm{E}}(t)$ corresponding to $x_{\mathrm{E}}(t)$
 is defined by
\beq
\bfq_{\mathrm{E}}(t)=\tilde{\bfq}(t; t).
\label{eq:envelope-def-van-der}
\eeq
It is now easy to show that $\bfq_{\mathrm{E}}(t)$ satisfies 
Eq. (\ref{eq:coupled-van-der-pol})
 up to the same order as $\tilde{\bfq}(t; t_0)$ does:
 In fact,  $ \forall t=t_0$
\beq
\frac{d\bfq _{\mathrm{E}}(t)}{d t}\Biggl\vert_{t= t_0}&=&
\frac{d\tilde{\bfq}(t; t_0)}{dt}\Biggl\vert _{t=t_0}
 + \frac{\d \tilde{\bfq}(t; t_0)}{\d t_0}\Biggl\vert_{t=t_0} \nonumber \\
 &=& \frac{d\tilde{\bfq}(t; t_0)}{dt}\Biggl\vert _{t=t_0} \nonumber \\
 &=&
{\bfF}(\tilde{\bfq}(t; t); \eps)+O(\eps^n) \nonumber \\
 &=&
{\bfF}({\bfq}_{\mathrm{E}}(t); \eps)+O(\eps^n).
\label{eq:van-der-proof}
\eeq
This completes the proof.
Here  Eq. (\ref{eq:rg-envelope-van-der}) and Eq. (\ref{eq:van-der-coupl-approx}) 
have been used together with the definition of ${\bfq}_{\mathrm{E}}(t)$, 
 Eq. (\ref{eq:envelope-def-van-der}). 
  It should be stressed that Eq. (\ref{eq:van-der-proof}) is valid
 uniformly  $\forall t$ \textit{i.e.},
in the global domain of $t$,
 in contrast to Eq. (\ref{eq:van-der-coupl-approx}) which is 
in a local domain  around $t=t_0$.

  We can summarize what we have done as follows:
  when there exist zero modes of the unperturbed operator,
  the higher-order corrections may give rise to secular terms,
  which are renormalized into
  the integral constants in the zeroth-order solution using the RG/envelope equation,
  and thereby, the would-be integral constants
  are lifted to dynamical  variables.

It will be found that the would-be integral constants corresponding to $A$ and $\phi$
exactly correspond to
  the hydrodynamic variables characterizing the local equilibrium distribution
function, such as the temperature $T$,
  chemical potential $\mu$,
  and flow velocity $u^{\mu}$ ($u^{\mu} \, u_{\mu} = 1$):
The equations (\ref{eq:amp-phase-eq})
 of the amplitude and phase which parametrize the solution also  
exactly correspond to the hydrodynamic equation governing the hydrodynamic
variables which\break
parametrize
the distribution function as the solution of the
Boltzmann equation.

\setcounter{equation}{0}
\section{Relativistic Boltzmann equation}
The relativistic Boltzmann equation (RBE) reads \cite{mic001,mic001-2}
\begin{eqnarray}
  \label{eq:1-001}
  p^\mu  \partial_\mu f_p(x) = C[f]_p(x),
\end{eqnarray}
where $f_p(x)$ denotes the one-particle distribution function
with $p^{\mu}$ being  the four-momentum of the on-shell particle, {\it i.e.},
$p^\mu p_\mu = p^2 = m^2$ and $p^0 > 0$.
$C[f]_p(x)$ in  r.h.s. denotes the collision integral 
\begin{eqnarray}
  \label{eq:1-002}
  C[f]_p(x) \equiv&& \frac{1}{2!}\sum_{p_1} \, \frac{1}{p_1^0}
  \sum_{p_2} \frac{1}{p_2^0} \sum_{p_3}  \frac{1}{p_3^0} 
  \omega(p, p_1|p_2, p_3)\nonumber \\
  && \times \Big( f_{p_2}(x) f_{p_3}(x) - f_p(x) f_{p_1}(x) \Big),
\end{eqnarray}
where $\omega(p \,,\, p_1|p_2 \,,\, p_3)$ denotes
the transition probability due to the microscopic two-particle interaction
with the symmetry property
\begin{eqnarray}
\omega(p, p_1|p_2, p_3) &=& \omega(p_2, p_3|p, p_1)\nonumber\\
&=& \omega(p_1, p|p_3, p_2) = \omega(p_3 , p_2|p_1 , p),
\end{eqnarray}
and the energy-momentum conservation
\begin{eqnarray}
  \omega(p \,,\, p_1|p_2 \,,\, p_3) \propto \delta^4(p + p_1 - p_2 - p_3).
\end{eqnarray}
To make explicit the correspondence 
to the general formulation of the reduction theory given in \cite{kuramoto,env006},
we treat the momentum as a discrete variable;
apart from such a formal reasoning, the summation
with respect to the momentum may be interpreted as the integration in practical use
 as
follows,
\begin{eqnarray}
  \sum_q \equiv \int\!\!d^3\Vec{q},
\label{eq:sum-integral}
\end{eqnarray}
with
$\Vec{q}$ being the spatial components of the four momentum $q^\mu$.

For an arbitrary vector $\varphi_p(x)$, 
the collision operator satisfies the following identity thanks to the 
above- \break mentioned symmetry property,
\begin{eqnarray}
  \label{eq:coll-symm}
  &&\hspace{-1cm}\sum_{p} \frac{1}{p^0} \varphi_p(x) C[f]_p(x)\nonumber\\
  &=& \frac{1}{2!}\frac{1}{4}\, \sum_{p,p_1\sim p_3} \, \frac{1}{p^0p^0_1p^0_2p^0_3} \, \omega(p , p_1|p_2 , p_3)\nonumber \\
  &&{}
  \times \big(\varphi_p(x)+\varphi_{p_1}(x)- \varphi_{p_2}(x) -\varphi_{p_3}(x)\big)
  \nonumber \\
  &&{}
  \times
  \big( f_{p_2}(x) \, f_{p_3}(x) - f_p(x) \, f_{p_1}(x) \big).
\end{eqnarray}
Substituting $(1,\,p^\mu)$ into $\varphi_p(x)$ in Eq. (\ref{eq:coll-symm}),
we find that $(1,\,p^\mu)$ are collision invariants satisfying
\begin{eqnarray}
  \label{eq:1-005}
  \sum_p \, \frac{1}{p^0} \, C[f]_p(x) = \sum_p \, \frac{1}{p^0} \, p^\mu \, C[f]_p(x) = 0,
\end{eqnarray}
due to the particle-number and energy-momentum conservation in the collision process,
respectively.
We note that
the function $\varphi_{0p}(x) \equiv a(x) + p^{\mu} \, b_{\mu}(x)$ 
is also a collision invariant
where $a(x)$ and $b_{\mu}(x)$ are arbitrary functions of $x$.
This form is, in fact, 
the most general form of a collision invariant \cite{mic001};
see \cite{mic001-2} for a proof.

Owing to the particle-number and energy- \break
momentum conservation in the collision process leading
to Eq. (\ref{eq:1-005}),
we have the balance equations
for the particle current $N^\mu(x)$ and the energy-momentum tensor $T^{\mu\nu}(x)$,
\begin{eqnarray}
  \label{eq:1-007}
  \partial_\mu N^\mu(x)
  \equiv
  \partial_\mu \Bigg[\sum_p \, \frac{1}{p^0} \, p^\mu \, f_p(x)\Bigg]
  &=& 0,\\
  \label{eq:1-008}
  \partial_\nu T^{\mu\nu}(x)
  \equiv
  \partial_\nu \Bigg[\sum_p \, \frac{1}{p^0} \, p^\mu \, p^\nu \, f_p(x)\Bigg]
  &=& 0,
\end{eqnarray}
respectively.
It should be noted that
any dynamical properties are not contained in these equations
unless the evolution of $f_p(x)$
has been obtained as a solution to Eq. (\ref{eq:1-001}).

In the Boltzmann theory, the entropy current may be defined \cite{mic001} by
\begin{eqnarray}
  \label{eq:1-009}
  S^\mu(x) \equiv - \sum_p \, \frac{1}{p^0} \, p^\mu \, f_p(x) \,
  \Big[ \ln \Big( (2\,\pi)^3 \, f_p(x) \Big) - 1 \Big],\nonumber\\
\end{eqnarray}
where the factor $(2\pi)^3$ is necessary owing 
to our convention (\ref{eq:sum-integral})\cite{mic001}.
The entropy current $S^\mu(x)$ satisfies
\begin{eqnarray}
  \label{eq:1-010}
  \partial_\mu S^\mu(x) =  - \sum_p \, \frac{1}{p^0} \, C[f]_p(x) \, \ln \Big( (2\,\pi)^3 \, f_p(x) \Big),
\end{eqnarray}
due to Eq. (\ref{eq:1-001}).
One sees that
$S^\mu(x)$ is conserved only if $\ln ((2\,\pi)^3 \, f_p(x))$ is a collision invariant,
{\it i.e.}, 
$\ln ((2\,\pi)^3 \, f_p(x)) = \varphi_{0p}(x) = a(x) + p^\mu \, b_\mu(x)$.
One thus finds \cite{mic001,mic001-2} that entropy-conserving distribution function may 
be parametrized as
\begin{eqnarray}
  \label{eq:Juettner}
  f_p(x)
  = \frac{1}{(2\pi)^{3}} \,
  \exp \Bigg[ \frac{\mu(x)
  - p^\mu \, u_\mu(x)}{T(x)} \Bigg]
  \equiv f^{\mathrm{eq}}_p(x),\nonumber\\
\end{eqnarray}
with $u^\mu(x) \, u_\mu(x) = 1$.
The function (\ref{eq:Juettner})
 is identified with the local equilibrium distribution function
called the J\"{u}ttner function \cite{juettner}, where
$T(x)$, $\mu(x)$, and $u^\mu(x)$ in Eq. (\ref{eq:Juettner})
are 
the local temperature,
chemical potential,
and flow velocity, respectively; see \cite{mic001-2} for a proof.
These five variables 
are called hydrodynamic variables.
Due to the energy-momentum conservation in the collision process,
we see that  the collision integral identically vanishes 
for the local equilibrium distribution $f^{\mathrm{eq}}_p(x)$ as
\begin{eqnarray}
  \label{eq:1-012}
  C[f^{\mathrm{eq}}]_p(x) = 0.
\end{eqnarray}

Some remarks are in order here.
In the proof \cite{mic001-2}, the Gibbs-Duhem relation as given by
\beq
d(s/n)=\frac{1}{T}\left(d(e/n) -\frac{p}{n^2}dn\right),
\label{eq:G-D}
\eeq
is taken for granted, where  $s$, $e$, $n$, and $p$ denote
the entropy, internal energy per volume, 
 particle density, and pressure in the equilibrium state,
respectively.
However,
Van and Biro \cite{van-and-biro2012}
have recently argued that the conventional Gibbs-Duhem relation (\ref{eq:G-D})
may be modified so as to contain the contribution from the thermal flow 
in the local equilibrium state of a relativistic system, 
and given 
a different interpretation for $T(x)$, $\mu(x)$, and $u^\mu(x)$
  in (\ref{eq:Juettner}); this modified definition of the local equilibrium
state, they claim,
 leads to the relativistic hydrodynamic equation in the particle frame
with the stable equilibrium state.
Although this is certainly an interesting possibility,
we will not follow this novel interpretation in this review: 
We shall make some comments on some related problem below.

\setcounter{equation}{0}
\section{Reduction to hydrodynamic equation}
Let us try to solve the RBE (\ref{eq:1-001})
in the hydrodynamic regime, and thereby derive the hydrodynamic equations 
governing the hydrodynamic variables.

\subsection{
  Relativistic Boltzmann equation
  in local rest frame of flow velocity
}
To make it explicit to solve the RBE in the hydrodynamic regime,
we first convert the RBE (\ref{eq:1-001}) into the following form
with the use of the
 {\em flow} velocity $u^{\mu}$ \, ($u^{\mu}u_{\mu}=1$) \cite{mic001}:
\begin{eqnarray}
  \label{eq:start}
  \frac{\partial}{\partial \tau} f_p(\tau,\, \sigma)
  &=& \frac{1}{p \cdot u} C[f]_p(\tau,\, \sigma)\nonumber \\
  &&{} - \varepsilon \, \frac{1}{p \cdot u}
  \, p \cdot \nabla f_p(\tau,\, \sigma),
\end{eqnarray}
where the new coordinate system $(\tau,\,\sigma^\mu)$
is defined as follows,
\begin{eqnarray}
  \label{eq:temporal-d}
  \frac{\partial}{\partial\tau} &=& u^\mu\,\partial_\mu \equiv D,\\
  \label{eq:spatial-d}
  \frac{\partial}{\partial\sigma_\mu} &=& (g^{\mu\nu} - u^\mu\,u^\nu)\,\partial_\nu \equiv \Delta^{\mu\nu} \, \partial_\nu \equiv \nabla^\mu.
\end{eqnarray}
We note that
$D$ and $\nabla^\mu$ are temporal and spatial differential operators familiar in
the  literature.
In Eq. (\ref{eq:start}),
the small parameter $\varepsilon$ is introduced 
as a measure of the non-uniformity of the fluid,
which may be identified with  
the Knudsen number;
$\varepsilon$ will be set back to unity 
in the final stage of the analysis.
In the present analysis based on the RG method, 
the perturbative expansion of the distribution function 
with respect to $\varepsilon$ is first performed
with the zeroth-order being the local equilibrium one; the dissipative effect
is taken into account as a deformation of the distribution function made by 
the spatial inhomogeneity as the perturbation.
Thus the above rewrite of the equation with $\varepsilon$ reflects a physical assumption
that only the spatial inhomogeneity is the origin of the dissipation.
It is noteworthy that
our RG method applied to the nonrelativistic Boltzmann equation
with the corresponding assumption successfully leads 
to the Navier-Stokes equation \cite{env007};
the present approach \cite{env009,Tsumura:2011cj}
is simply a relativistic generalization of the nonrelativistic case.

Here we make a comment on the possibility of a rewrite of 
the RBE (\ref{eq:1-001}) with use of a different
time-like four vector in place of the flow
velocity $u^{\mu}$. 
In other words,
we examine whether
Eq. (\ref{eq:start}) with  $u^{\mu}$ being identified with the flow velocity 
is a unique rewrite of the RBE (\ref{eq:1-001}) in a covariant manner.
We argue that it is the case on the basis of a physical ground.

We first introduce a generic time-like four vector $\Ren{a}^\mu$ with $\Ren{a}^2\,>\,0$,
and call it the \textit{macroscopic-frame vector}, following \cite{env009,Tsumura:2012gq}.
Without a loss of generality, the generic vector of Lorentz covariance
takes the form,
\begin{eqnarray}
\label{eq:general-a}
\Ren{a}^\mu = A_1\,u^\mu + A_2\,\partial^\mu T
+ A_3\,\partial^\mu \mu + A_4\,u^\nu\,\partial_\nu u^\mu,
\end{eqnarray}
since $u^{\mu}$ and $\d^{\mu}$ are the only available Lorentz vectors at hand.
Here, $A_1$, $A_2$, $A_3$, and $A_4$ are arbitrary functions of 
the temperature $T$ and the chemical potential $\mu$;
$A_i = A_i(T,\,\mu)$ for $i=1,\,2,\,3,\,4$.
Owing to the identity 
\begin{eqnarray}
\d^{\mu}=u^{\mu}u^{\nu}\d_{\nu}+(g^{\mu\nu}-u^{\mu}u^{\nu})\d_{\nu} = u^{\mu}D+\nabla^{\mu},
\end{eqnarray}
where $D$ and $\nabla^\mu$ have been defined in Eq.'s (\ref{eq:temporal-d}) and (\ref{eq:spatial-d}),
Eq.(\ref{eq:general-a}) is  rewritten as
\begin{eqnarray}
\label{eq:reduced-a-mu}
\Ren{a}^\mu &=&\big(A_1 + A_2  DT + A_3 \, D\mu\big) \, u^\mu 
\nonumber\\
&&{}+ A_2 \, \nabla^\mu T + A_3 \, \nabla^\mu\mu
+ A_4\,Du^\mu
\nonumber \\
&\equiv& C_t(T,\,\mu) \, u^\mu+A_2 \, \nabla^\mu T + A_3 \, \nabla^\mu\mu + A_4 \, Du^\mu,\nonumber\\
\end{eqnarray}
with $C_t(T,\,\mu) = A_1 + A_2 \, DT + A_3 \, D\mu$. 
The relative magnitudes 
of $C_t$ and $A_{2,3,4}$ are only constrained by the inequality $\Ren{a}^2>0$ in the present
stage. However, it should be emphasized that the space-like terms with the coefficients  
$A_{2,3,4}$ are all derivative terms, which are supposed to be small in the hydrodynamic regime even 
in the dissipative regime if the dynamics is governed the hydrodynamics at all.

By replacing $u^{\mu}$ by $\Ren{a}^\mu$ given by (\ref{eq:reduced-a-mu}),
we have the generic coordinate system
$(\tilde{\tau},\,\tilde{\sigma}^\mu)$ as defined by
\begin{eqnarray}
  \frac{\partial}{\partial\tilde{\tau}} &\equiv& \frac{\Ren{a}^\mu}{\Ren{a}^2}\,\partial_\mu,\\
  \frac{\partial}{\partial\tilde{\sigma}_\mu} &\equiv& \Big( g^{\mu\nu} - \frac{\Ren{a}^\mu\,\Ren{a}^\nu}{\Ren{a}^2} \Big) \, \partial_\nu.
\end{eqnarray}
With this coordinate system, the RBE is rewritten 
as 
\begin{eqnarray}
  \frac{\partial}{\partial \tilde{\tau}} f_p(\tilde{\tau},\, \tilde{\sigma})
  &=& \frac{1}{p \cdot \Ren{a}} C[f]_p(\tilde{\tau},\, \tilde{\sigma})\nonumber \\
  &&{} - \varepsilon \, \frac{1}{p \cdot \Ren{a}}
  \, p^\mu \, \frac{\partial}{\partial\tilde{\sigma}^\mu} f_p(\tilde{\tau},\, \tilde{\sigma}),
\end{eqnarray}
where the $\varepsilon$ is again multiplied to $\partial/\partial\tilde{\sigma}^\mu$
as was done in Eq. (\ref{eq:start}) 
where
it is supposed that 
only the spatial inhomogeneity is the origin of the dissipation.
Then the space-like terms with the coefficients $A_2$ and $A_3$ in $\Ren{a}^\mu$ 
are of higher order with respect to $\varepsilon$
and should be ignored in this set up. Furthermore, 
since we start with a stationary solution with vanishing time-dependence in the RG approach,
the term with $A_4$ should be also ignored. Thus we have 
\begin{eqnarray}
  \Ren{a}^\mu = C_t(T,\,\mu)\,u^\mu \equiv b \, u^\mu,
\end{eqnarray}
and accordingly
\begin{eqnarray}
  \frac{\partial}{\partial\tilde{\sigma}_\mu} = \nabla^\mu.
\end{eqnarray}
If we naturally require that 
$b$ should be independent of the momentum $p^\mu$,
it is easy to show \cite{next} that
the ``normalization'' factor $b$ can be made unity without loss of generality,
in conformity of the natural choice \cite{mic001,mic001-2}
$\Ren{a}^\mu=u^\mu$.

It is remarkable 
that this natural choice
uniquely leads to the hydrodynamic equation in the energy (Landau-Lifshitz) frame, as will be
shown and discussed later \cite{env009,Tsumura:2012gq}.
Conversely,
a  choice of $b$ different from
unity with a momentum dependence
could lead to various hydrodynamic equations other than that of Landau and Lifshitz,
including the one in the particle frame for viscous fluids 
as was shown by the present authors \cite{env009,Tsumura:2012gq}. 
However, it is worth emphasizing  that 
the particle frame can be only realized when
$b$ has a peculiar momentum dependence
such as $b = m/(p\cdot u)$ ($\Ren{a}^\mu = (m/(p\cdot u))\,u^\mu$)
\cite{env009,Tsumura:2012gq}.
In retrospect, however,
the possible momentum dependence of $b$ can not be legitimate for 
$\Ren{a}^\mu $ to play a macroscopic-frame vector,
because it means that
the covariant and macroscopic space-time in the particle frame is defined 
for a respective particle state with a definite energy-momentum,
which is certainly unnatural and  lead
 to a trouble in a physical interpretation \cite{next}.
Thus,
we naturally require that 
$b$ is independent of the momentum $p^\mu$ and hence
$\Ren{a}^\mu = u^\mu$.

 \subsection{
Hydrodynamics from
 relativistic Boltzmann equation by renormalization-group method
  }
  \label{sec:2}
  
Applying the perturbation theory to Eq. (\ref{eq:start}),
  we derive the relativistic dissipative hydrodynamic equation
  as the infrared asymptotic dynamics
  of the RBE  by the RG method \cite{env001,env006,env009,Tsumura:2011cj}.

In this approach,
we first try to obtain the perturbative solution $\tilde{f}_p$ 
  to Eq. (\ref{eq:start})
  around the arbitrary initial time $\tau = \tau_0$
  with the initial value $f_p(\tau_0 , \sigma)$;
\begin{eqnarray}
 \tilde{f}_p(\tau = \tau_0 , \sigma ; \tau_0) = f_p(\tau_0 , \sigma).
  \end{eqnarray}
Note that the solution depends on the initial time $\tau_0$ at which
$\tilde{f}_p(\tau = \tau_0 , \sigma ; \tau_0)$
 is supposed to be on an exact solution.
We expand the initial value as well as the solution
  with respect to $\varepsilon$ as follows:
\beq
\tilde{f}_p(\tau, \sigma ; \tau_0)
    &=& \tilde{f}_p^{(0)}(\tau , \sigma ; \tau_0)
    + \varepsilon  \tilde{f}_p^{(1)}(\tau , \sigma ; \tau_0)\nonumber \\
&{}&    + \varepsilon^2  \tilde{f}_p^{(2)}(\tau , \sigma ; \tau_0)
    + \cdots, \\
{f}_p(\tau_0, \sigma)
    &=& {f}_p^{(0)}(\tau_0 , \sigma)
    + \varepsilon  {f}_p^{(1)}(\tau_0, \sigma) \nonumber \\
  &{}&    + \varepsilon^2  \tilde{f}_p^{(2)}(\tau , \sigma ; \tau_0)
    + \cdots.
\eeq
    
The zeroth-order equation reads
  \begin{eqnarray}
    \label{eq:1-025}
    \frac{\partial}{\partial \tau} \tilde{f}^{(0)}_p(\tau \,,\, \sigma \,;\, \tau_0)
    = \frac{1}{p \cdot u} \,
    C[\tilde{f}^{(0)}]_p(\tau \,,\, \sigma \,;\, \tau_0).
  \end{eqnarray}
  Since we are looking for the slow motion
to be realized asymptotically when $\tau \rightarrow \infty$,
  we take the stationary solution satisfying
\begin{eqnarray}
  \label{eq:1-026}
\frac{\partial}{\partial \tau}\tilde{f}_p^{(0)}(\tau , \sigma ; \tau_0) = 0,
\eeq
implying that
$C[\tilde{f}^{(0)}]_p(\tau , \sigma ; \tau_0) = 0$\, $\forall\, \sigma$,
which is solved by 
  a local equilibrium distribution function, \textit{i.e.}, the J\"uttner distribution
function,
   \begin{eqnarray}
    \label{eq:1-028}
    \tilde{f}_p^{(0)}(\tau , \sigma ; \tau_0)
    &=& \frac{1}{(2\pi)^{3}} \,
    \exp \Bigg[ \frac{\mu(\sigma ; \tau_0)
        - p^\mu \, u_\mu(\sigma ; \tau_0)}{T(\sigma ; \tau_0)} \Bigg] \nonumber \\
    &\equiv& f^{\mathrm{eq}}_p(\sigma ; \tau_0),
  \end{eqnarray}
  with $u^\mu(\sigma ; \tau_0) \, u_\mu(\sigma ; \tau_0) = 1$.
 Here the  would-be integration constants
  $T(\sigma \,;\, \tau_0)$, $\mu(\sigma \,;\, \tau_0)$, and $u_\mu(\sigma \,;\, \tau_0)$
  are independent of $\tau$ but may depend on $\tau_0$ as well as $\sigma$.
  
Now that the zero-th order solution is given, the first-order equation reads
  \begin{eqnarray}
    \label{eq:1-030}
    \frac{\partial}{\partial \tau} \tilde{f}_p^{(1)}(\tau)
    = \sum_q \, A_{pq} \, \tilde{f}_q^{(1)}(\tau) + F_p,
  \end{eqnarray}
with
\beq
F_p\equiv -  \frac{1}{p \cdot u} \, p \cdot \nabla f^{\mathrm{eq}}_p,
\eeq
where $A_{pq}$ denotes a matrix element of the linearized collision operator 
$A$; \textit{i.e.}, 
  \begin{eqnarray}
    \label{eq:1-031}
    (A)_{pq}=A_{pq} \equiv \frac{1}{p \cdot u} \,
    \frac{\partial}{\partial f_q} C[f]_p \, \Bigg|_{f =  f^{\mathrm{eq}}}.
  \end{eqnarray}

 
Let us examine 
 the spectral properties of $A$;
for which, it is found convenient
to convert $A$ to another linear operator,
\beq
 L \equiv (f^{\mathrm{eq}})^{-1} \, A \, f^{\mathrm{eq}},
\eeq
  with the diagonal matrix
  $(f^\mathrm{eq})_{pq} \equiv f^\mathrm{eq}_p \, \delta_{pq}$.
Next we define an inner product
  between arbitrary nonzero vectors $\varphi$ and $\psi$ by
  \begin{eqnarray}
    \label{eq:1-034}
    \langle  \, \varphi \,,\, \psi \, \rangle
    \equiv \sum_{p} \, \frac{1}{p^0} \, (p \cdot u) \,
    f^{\mathrm{eq}}_p \, \varphi_p \, \psi_p,
  \end{eqnarray}
which satisfies the positive-definiteness of the norm as
  \begin{eqnarray}
    \label{eq:1-035}
    \langle  \varphi\,,\,\varphi \, \rangle =
    \sum_{p} \, \frac{1}{p^0} \, (p \cdot u) \,
    f^{\mathrm{eq}}_p \, (\varphi_p)^2 >  0
  \end{eqnarray}
for $\varphi_p \ne 0$,
  since both $p^{\mu}$ and $u^{\mu}$ are time-like vectors with $p^{0}>0$. 

Then it can be shown \cite{env009,Tsumura:2011cj} 
that the linearized collision operator $L$ has remarkable properties that 
it is semi-negative definite and has five zero modes  given by
  \begin{eqnarray}
    \label{eq:1-039}
    \varphi_{0p}^\alpha \equiv \left\{
    \begin{array}{ll}
      \displaystyle{p^\mu} & \displaystyle{\mathrm{for}\,\,\,\alpha = \mu}, \\[2mm]
      \displaystyle{1\times m}     & \displaystyle{\mathrm{for}\,\,\,\alpha = 4}.
    \end{array}
    \right.
  \end{eqnarray}
The functional subspace spanned by the five zero modes is called the P$_0$ space and
the projection operator to it is denoted by $P_0$; 
\begin{eqnarray}
    \label{eq:1-041}
    \big[ P_0 \, \psi \big]_p &\equiv&
    \varphi_{0p}^\alpha \, \eta^{-1}_{0\alpha\beta} \,
    \langle \, \varphi_0^\beta \,,\, \psi \, \rangle,
  \end{eqnarray}
  where
  $\eta^{-1}_{0\alpha\beta}$ is the inverse matrix of
  the the P${}_0$-space metric matrix $\eta_0^{\alpha\beta}$
  defined by
  \begin{eqnarray}
    \label{eq:1-043}
    \eta_0^{\alpha\beta} \equiv \langle \, \varphi_0^\alpha \,,\, \varphi_0^\beta \, \rangle.
  \end{eqnarray}
We also call the complement to P$_0$ the Q$_0$ space and introduce $Q_0 \equiv 1 - P_0$.
In the following, we also use the modified projection operators 
defined by
\beq
\bar{P}_0=f^{\rm eq}P_0(f^{\rm eq})^{-1}, \quad
\bar{Q}_0=f^{\rm eq}Q_0(f^{\rm eq})^{-1},
\eeq
which means, for example, 
\beq
\big[\bar{P}_0\, \psi \big]_p =
    f^{\mathrm{eq}}_p \, \varphi_{0p}^\alpha \, \eta^{-1}_{0\alpha\beta} \,
    \langle \, \varphi_0^\beta \,,\, (f^{\rm eq})^{-1}\psi \, \rangle.
  \end{eqnarray} 

Then  the perturbative solution up to the second order 
reads
\beq
\tilde{f}_p(\tau , \sigma ; \tau_0)
   & =& \tilde{f}^{(0)}_p(\tau , \sigma ; \tau_0)
    + \varepsilon  \tilde{f}^{(1)}_p(\tau,\, \sigma ; \tau_0)\nonumber \\
  &{}&
   + \varepsilon^2 \tilde{f}^{(2)}_p(\tau , \sigma ; \tau_0)
    + O(\varepsilon^3),
\eeq
where
$\tilde{f}^{(1)}(\tau, \sigma ; \tau_0) = (\tau - \tau_0) \bar{P}_0  F - A^{-1}  \bar{Q}_0  F$
and a lengthy formula for
$\tilde{f}^{(2)}(\tau, \sigma ; \tau_0 )$, which we do not write down for the sake of space;
 see \cite{Tsumura:2011cj} for the details.

  We remark  that this solution contains
   secular terms, which apparently invalidates the perturbative expansion
  for $\tau$ away from the initial time $\tau_0$.
  We can, however, utilize the secular terms to obtain
  an asymptotic solution valid in a global domain \cite{env001,env006}.
  Indeed   we have a family of curves
  $\tilde{f}_p(\tau,\, \sigma;\, \tau_0)$
  parameterized with $\tau_0$:
  They are all on the exact solution
  $f_p(\sigma \,;\, \tau)$ at $\tau = \tau_0$ up to $O(\varepsilon^3)$,
  although only valid  for $\tau$ near $\tau_0$ locally.
  Then,  the {\em envelope curve} of the family of curves,
  which is in contact with each local solution at $\tau = \tau_0$, will
  give a global solution in our asymptotic situation, which is shown to 
be the case \cite{env001,env006}.
  According to the classical theory of envelopes,
  the envelope that is in contact with any curve in the family
  at $\tau = \tau_0$ is obtained \cite{env001} by
  \begin{eqnarray}
    \label{eq:1-058}
    \frac{d}{d\tau_0}
    \tilde{f}_p(\tau , \sigma ; \tau_0) \Bigg|_{\tau_0 = \tau} = 0.
  \end{eqnarray}
The derivative  w.r.t. $\tau_0$ hits the hydrodynamic variables, and hence we have 
the evolution equation of them that is identified with the hydrodynamic 
equation \cite{env007,env009}. We also note that the invariant manifold 
which corresponds to the hydrodynamics
 in the functional space of the distribution function is explicitly obtained as 
an envelope function \cite{env009,Tsumura:2011cj}:
$f_{\mathrm{E}p}(\tau,\sigma) =    \tilde{f}_p(\tau,\,\sigma \,;\, \tau_0 = \tau)$,
 the explicit form of which is referred to \cite{env009,Tsumura:2011cj}.
We note that this solution is valid in a global domain 
of time in the asymptotic region \cite{Tsumura:2011cj}.

  Putting back $\varepsilon$ to $1$,
Eq. (\ref{eq:1-058}) is  reduced to the following form in this
approximation,
  \begin{eqnarray}
    \label{eq:1-061}
    \sum_{p}  \frac{1}{p^0}  \varphi_{0p}^\alpha 
    \Bigg[ (p \cdot u)  \frac{\partial}{\partial \tau}
      + p \cdot \nabla \Bigg]
    ( f^{\mathrm{eq}}_p +  \delta f^{(1)}_p ) = 0.
  \end{eqnarray}
where $\delta f^{(1)}_p$ denotes
 the first-order correction to the distribution function
\beq
\delta f^{(1)}_p \equiv -[ A^{-1}  \bar{Q}_0  F ]_p.
\label{eq:phi-bar}
\eeq
  If one uses the identity
$(p \cdot u) \, \partial/\partial \tau
    + p \cdot \nabla = p^\mu\,\partial_\mu$,
Eq. (\ref{eq:1-061}) is found to have the following form
  \begin{eqnarray}
    \label{eq:1-064}
    \partial_\mu T^{\mu\nu}_{\mathrm{1st}} = 0,\quad
    \partial_\mu N^{\mu}_{\mathrm{1st}} = 0.
  \end{eqnarray}
  with
$T^{\mu\nu}_{\mathrm{1st}} =T^{(0)\mu\nu}+\delta \, T^{\mu\nu}_{\mathrm{1st}}$ and
$N^{\mu}_{\mathrm{1st}}=N^{(0)\mu}+\delta\, N^{\mu}_{\mathrm{1st}}$.
Here, 
  \begin{eqnarray}
    \label{eq:1-065-ideal}
    T^{(0)\mu\nu}
    &\equiv& \sum_{p} \frac{1}{p^0}  p^\mu p^{\nu} f^{\mathrm{eq}}_p
    = e\,u^{\mu}\,u^{\nu}-p\,\Delta^{\mu\nu},\\
    N^{(0)\mu} &\equiv& \sum_{p}  \frac{1}{p^0}  p^\mu f^{\mathrm{eq}}_p = n\,u^{\mu},
  \end{eqnarray}
with $e$, $p$, and $n$ being
the internal energy, pressure, and particle-number density
for the relativistic ideal gas, respectively,
while the dissipative parts are given 
as a deviation of the local equilibrium distribution function
  \begin{eqnarray}
    \label{eq:1-065}
   \delta T^{\mu\nu}_{\mathrm{1st}}
    &\equiv& \sum_{p}  \frac{1}{p^0}  p^\mu p^{\nu}\delta f^{(1)}_p , \\
\label{eq:delta-N}   
 \delta N_{\mathrm{1st}}
    &\equiv& \sum_{p}  \frac{1}{p^0}  p^\mu \delta f^{(1)}_p.
  \end{eqnarray}
As is well known,
the local equilibrium distribution function as given by (\ref{eq:1-028}) only gives
the relativistic Euler equation without dissipation.

\subsection{Possible uniqueness of Landau-Lifshitz frame}
In this subsection, 
we present the explicit form of  the dissipative 
parts
$\delta T^{\mu\nu}_{\mathrm{1st}}$ and $\delta N^{\mu}_{\mathrm{1st}}$
and discuss their
properties.
An  evaluation of Eq. (\ref{eq:1-065}) together with (\ref{eq:phi-bar})
gives \cite{env009,Tsumura:2011cj},
  \begin{eqnarray}
    \label{eq:2-068}
  \delta  T^{\mu\nu}_{\mathrm{1st}} &=&  \zeta \,\Delta^{\mu\nu}\, \nabla\cdot u
    + 2 \, \eta \, \Delta^{\mu\nu\rho\sigma} \, \nabla_\rho u_\sigma,\\
    \label{eq:2-069}
  \delta N^\mu_{\mathrm{1st}} &=& \lambda \, \frac{1}{\hat{h}^2} \, \nabla^\mu\frac{\mu}{T},
  \end{eqnarray}
respectively,
with 
$\Delta^{\mu\nu\rho\sigma} \equiv 
1/2 \cdot (\Delta^{\mu\rho}\Delta^{\nu\sigma} + \Delta^{\mu\sigma}\Delta^{\nu\rho}
- 2/3 \cdot \Delta^{\mu\nu}\Delta^{\rho\sigma})$.
Here,
$\hat{h}$ denotes
the reduced enthalpy per particle.
The bulk and shear viscosities and the thermal conductivity are
denoted by $\zeta$, $\eta$ and $\lambda$, respectively. 
 It is clear that these formulas completely agree with those 
 proposed by Landau and Lifshitz \cite{hen002}.
  Indeed,
  the respective dissipative parts
$\delta T^{\mu\nu}_{\mathrm{1st}}$ and $\delta N^{\mu}_{\mathrm{1st}}$
  in Eq.'s (\ref{eq:2-068}) and (\ref{eq:2-069}) meet Landau and Lifshitz's ansatz
\beq
\label{eq:delta-e}
\delta e &\equiv& u_\mu  \,\delta T^{\mu\nu}_{\mathrm{1st}} \,u_\nu = 0, \\
\label{eq:delta-n}
\delta n &\equiv& u_\mu \, \delta N^\mu_{\mathrm{1st}} = 0, \\
\label{eq:delta-Q}
Q_\mu &\equiv& \Delta_{\mu\nu} \,\delta T^{\nu\rho}_{\mathrm{1st}} \,u_\rho = 0.
\eeq
Thus we find that the frame on which the flow velocity is 
defined  inevitably becomes 
the Landau-Lifshitz (energy) frame, if 
the hydrodynamics is to be consistent with the underlying
relativistic Boltzmann equation
\footnote{The uniqueness of the energy frame for the 
relativistic hydrodynamics is recently argued also in a different context \cite{minami-hidaka}.}.

Let us  see the above fact in the level of the distribution function.
We first note that Eq. (\ref{eq:1-065}) and (\ref{eq:delta-N}) can be rewritten as
\beq
    \label{eq:1-065-2}
   \delta T^{\mu\nu}_{\mathrm{1st}}
    &=& \sum_{p}  \frac{1}{p^0}  p^\mu p^{\nu}f^{\mathrm{eq}}_p\bar{\phi}_p , \\
\label{eq:delta-N-2}   
 \delta N_{\mathrm{1st}}
    & = & \sum_{p}  \frac{1}{p^0}  p^\mu f^{\mathrm{eq}}_p\bar{\phi}_p,
  \end{eqnarray}
   with
\beq
\bar{\phi}_p = -\big[L^{-1}Q_0(f^{\mathrm{eq}})^{-1}F\big]_p,
\eeq
which belongs to the Q$_0$ space and thus orthogonal to the zero modes,
  \begin{eqnarray}
    \label{eq:4-005}
    \langle \varphi^\alpha_0 \,,\, \bar{\phi}  \rangle = 0
    \,\,\,\mathrm{for}\,\,\,\alpha = 0,\,1,\,2,\,3,\,4.
  \end{eqnarray}
  Here, the inner product is defined by Eq. (\ref{eq:1-034}).
  Then, Eq. (\ref{eq:4-005}) with $\alpha =\mu$ is reduced to
  \begin{eqnarray}
    \label{eq:4-006}
   0&=& \sum_p  \frac{1}{p^0} (p \cdot u)  f^\mathrm{eq}_p 
    p^{\mu}\bar{\phi}_p 
 = u_{\nu}\sum_p  \frac{1}{p^0} p^{\nu}p^{\mu} f^\mathrm{eq}_p \bar{\phi}_p \nonumber \\
 &=& u_{\nu}\,\delta T^{\mu\nu}_{\mathrm{1st}}.
  \end{eqnarray}
Similarly, Eq. (\ref{eq:4-005}) with $\alpha =4$ is reduced to
$u_\mu \, \delta N^\mu_{\mathrm{1st}} = 0$.
  Thus, one can readily see that these equations coincide with
  Landau and Lifshitz's ansatz:
We remark that Eq. (\ref{eq:4-006}) implies the following two equations,
  $\delta e \equiv u_\mu \, u_\nu \, \delta T^{\mu\nu}_{\mathrm{1st}}$ $=0$ and 
$Q_{\rho} \equiv \Delta_{\rho\mu} \, u_{\nu} \, \delta T^{\mu\nu}_{\mathrm{1st}} = 0$, 
which are nothing but the matching conditions \cite{mic001} 
imposed to select the energy frame
 in all the other existing approaches based on the Boltzmann equation.
In other words, we have given the foundation to the matching conditions \cite{mic001} 
for the energy frame.

We can present an intuitive picture of
why the energy frame is uniquely selected in this method.
As shown in \cite{Tsumura:2012gq},
the physical quantity transported by each particle governed by the RBE (\ref{eq:start})
can be identified with $(p\cdot u)$.
To clarify the physical meaning of $(p\cdot u)$,
we take the non-relativistic limit of this quantity:
\begin{eqnarray}
  (p\cdot u) \sim m + \frac{m}{2}\,\Big|\frac{\Vec{p}}{p^0} - \Vec{u}\Big|^2,
\end{eqnarray}
where $u^\mu = (u^0,\,\Vec{u})$ and $p^\mu = (p^0,\,\Vec{p})$.
This equation shows that
$(p\cdot u)$ can be interpreted
as the kinetic energy of the fluid component measured in the rest frame of $u^\mu$.
Thus, it is natural that
the resultant equation becomes the one
in the energy frame adopted by Landau and Lifshitz.

A remark is in order here.
The uniqueness of the energy frame comes from
the two natural conditions used in the derivation, \textit{i.e.}, 
the identification of the time-like vector $u^{\mu}$ in the J\"uttner distribution function (\ref{eq:1-028})
with the flow velocity
and the physical assumption that
the dissipative effect comes from only the spatial inhomogeneity.
If one of these conditions were to be challenged, as claimed in 
\cite{van-and-biro2012}, for instance,
the uniqueness of the energy frame could be violated.
It is clear that further studies are needed for establishing the 
uniqueness of the energy frame in the relativistic hydrodynamics for viscous fluids.

\subsection{Transport coefficients}

Since our theory starts from a microscopic theory as  statistical mechanics,
we have the microscopic expressions for 
the transport coefficients appearing in the hydrodynamic tensor(\ref{eq:2-068}) and current 
(\ref{eq:2-069}), as follows:
\beq
\label{eq:trans-coef-1st.1}
\zeta &=& - \frac{1}{T} \langle  \tilde{\Pi},L^{-1}\tilde{\Pi} \rangle,\\
\label{eq:trans-coef-1st.2}
\lambda &=& \frac{1}{3 T^2} \langle  \tilde{J}^\mu, L^{-1}\tilde{J}_\mu \rangle, \\
\label{eq:trans-coef-1st.3}
\eta &=& - \frac{1}{10 T}  \langle \tilde{\pi}^{\mu\nu}, L^{-1}\tilde{\pi}_{\mu\nu} \rangle.
\eeq
  Here, we have introduced the following microscopic thermal forces
  $(\tilde{\Pi}_p,\,\tilde{J}^\mu_p,\,\tilde{\pi}^{\mu\nu}_p) \equiv
 (\Pi_p,\,J^\mu_p,\,\pi^{\mu\nu}_p)/{(p\cdot u)}$, with
\beq
\Pi_p &\equiv& \Big({4}/{3} - \gamma \Big) (p \cdot u)^2
    + \Big( (\gamma - 1)  T \hat{h} - \gamma  T \Big)  (p \cdot u)\nonumber \\
    &&{} - {1}/{3}\cdot m^2,\\
J^\mu_p &\equiv& - ( (p \cdot u) - T  \hat{h} ) \Delta^{\mu\nu}  p_\nu,\\
\pi^{\mu\nu}_p &\equiv& \Delta^{\mu\nu\rho\sigma} \, p_\rho \,  p_\sigma.
\eeq
Here, $\gamma$ denotes
the ratio of the constant pressure and volume heat capacities.
We note that
the microscopic expressions for 
the transport coefficients (\ref{eq:trans-coef-1st.1})-(\ref{eq:trans-coef-1st.3})
are in agreement with those given by the Chapman-Enskog method \cite{mic001}.

It is noteworthy that
the transport coefficients can be
rewritten in the Green-Kubo formula \cite{text}.
With the use of  the ``time-dependent thermal force'' defined by
\begin{eqnarray}
\tilde{\Pi}_p(s)\equiv  \sum_q \left[ e^{sL} \right]_{pq}\tilde{\Pi}_q
\end{eqnarray}
and so on,  the relaxation functions are given by the time-correlators 
\beq
    \label{eq:1-037}
R_\zeta(s) &\equiv& \frac{1}{T} \, \langle \, \tilde{\Pi}(0)\,,\,\tilde{\Pi}(s) \,
    \rangle,
\eeq
and so on for $R_\lambda(s)$ and $R_\eta(s)$ with obvious modifications to $R_{\zeta}(s)$.
Then the transport coefficients given
in Eq.'s (\ref{eq:trans-coef-1st.1})-(\ref{eq:trans-coef-1st.3}) are rewritten as
follows \cite{env009,Tsumura:2011cj},
  \begin{eqnarray}
    \zeta = \int_0^\infty\!\!ds\,R_\zeta(s),\,\,
    \lambda = \int_0^\infty\!\!ds\,R_\lambda(s),\,\,
    \eta = \int_0^\infty\!\!ds\,R_\eta(s).\nonumber \\
 \end{eqnarray}

\setcounter{equation}{0}
  \section{
Generic stability
of relativistic hydrodynamic
    equation in energy frame
  }

  In this section, we shall provide a proof \cite{Tsumura:2011cj}
  that
generic constant
solutions of
  the relativistic dissipative hydrodynamic equation 
 in the energy frame 
  is stable against a small perturbation \cite{env010},
  on account of the positive definiteness of
  the inner product as shown in Eq. (\ref{eq:1-035}).

For this purpose,
we first note that $F_p$ is reduced to
 \begin{eqnarray}
    \label{eq:1-072-0}
    F_p 
    = - f^\mathrm{eq}_p\, \frac{1}{p\cdot u}\, p^\mu\,\varphi^\alpha_{0p}\, \nabla_\mu X_{\alpha},
  \end{eqnarray}
with $\varphi^\alpha_{0p}$ being the zero modes defined in (\ref{eq:1-039}) and 
  \begin{eqnarray}
    \label{eq:1-073-2}
    X_{\alpha} \equiv \left\{
    \begin{array}{ll}
      \displaystyle{- u_\nu / T}
      &
      \displaystyle{\mathrm{for}\,\,\,\alpha = \nu}, \\ [2mm]
      \displaystyle{m^{-1} \, \mu /T}
      &
      \displaystyle{\mathrm{for}\,\,\,\alpha = 4}.
    \end{array}
    \right.
  \end{eqnarray}
Then Eq. (\ref{eq:1-061}) is rewritten 
in the following form,
  \begin{eqnarray}
    \label{eq:1-078}
    &&\sum_{p}  \frac{1}{p^0}\varphi_{0p}^\alpha   
    \Bigg[ (p \cdot u) \frac{\partial}{\partial \tau}
      + p \cdot \nabla \Bigg]
    \Bigg[ f^{\mathrm{eq}}_p  \big(1 + \nonumber\\
 && [ L^{-1}  \varphi^{\nu\beta}_1 ]_p
 \nabla_{\nu}{X}_{\beta} \big) \Bigg] = 0,
  \end{eqnarray}
where
 \begin{eqnarray}
    \label{eq:1-077}
    \varphi^{\mu\alpha}_{1p} \equiv \big[ Q_0 \,
      \tilde{\varphi}^{\mu\alpha}_1 \big]_p,
  \end{eqnarray}
with 
$\tilde{\varphi}^{\mu\alpha}_{1p} \equiv p^\mu\varphi^\alpha_{0p}/(p\cdot u)$.
 
  Now, a
generic constant
solution means that
  it describes a system having a finite homogeneous flow
  with a constant temperature and a constant chemical potential, as follows:
  \begin{eqnarray}
    T(\sigma;\tau) = T_0,\,\,
    \mu(\sigma;\tau) = \mu_0,\,\,
    u_\mu(\sigma;\tau) = u_{0\mu},
   \label{eq:5-001}
   \end{eqnarray}
  where $T_0$, $\mu_0$, and $u_{0\mu}$ are constant.
  We note that
these
states include the
  thermal equilibrium state as a special case.
  
  We shall show the linear stability of the
constant
solution
  of the relativistic dissipative hydrodynamic equation in the energy frame.
  We represent $T$, $\mu$, and $u_\mu$ around the
constant
solution as follows:
  \begin{eqnarray}
    \label{eq:5-004}
    T(\sigma\,;\,\tau) &=& T_0 + \delta T(\sigma\,;\,\tau),\\
    \label{eq:5-005}
    \mu(\sigma\,;\,\tau) &=& \mu_0 + \delta\mu(\sigma\,;\,\tau),\\
    \label{eq:5-006}
    u_\mu(\sigma\,;\,\tau) &=& u_{0\mu} + \delta u_\mu(\sigma\,;\,\tau),
  \end{eqnarray}
where the deviations 
$\delta T$, $\delta \mu$, and $\delta u_\mu$ are assumed to so small
that terms in the second or higher orders of them
   can be neglected.
  Instead of these six variables which  not independent of each other
  because $\delta u_\mu \, u^\mu_0 = 0$,
  we use the following five independent variables,
  \begin{eqnarray}
    \label{eq:5-007}
    \delta X_{\alpha} &\equiv& \left\{
    \begin{array}{ll}
      \displaystyle{
        - \delta \left(\frac{u_\mu}{T}\right) = - \frac{\delta u_\mu}{T_0} + 
       \delta T \frac{u_{0\mu}}{T^2_0}
      }
      & \displaystyle{\mathrm{for}\,\,\alpha = \mu}, \\[2mm]
      \displaystyle{
        m^{-1} \, \delta \left(\frac{\mu}{T}\right) = m^{-1} \, (\frac{\delta \mu}{T_0} - 
        \delta T \frac{\mu_0}{T^2_0})
      }
      & \displaystyle{\mathrm{for}\,\,\alpha = 4}.
    \end{array}
    \right.\nonumber\\
  \end{eqnarray}
  Substituting Eq. (\ref{eq:5-007}) into Eq. (\ref{eq:1-078}) and with 
some manipulation,
  we obtain the linearized equation governing $\delta X_{\alpha}$ as
  \begin{eqnarray}
    \label{eq:5-008}
    &&\Big( \langle \varphi_{0}^\alpha ,  \varphi^\beta_0 \rangle
    + \langle \varphi_{0}^\alpha, L^{-1} \varphi^{\nu\beta}_1
    \rangle \nabla_\nu\Big) 
    \frac{\partial}{\partial \tau}
    \delta X_\beta\nonumber\\
    &&\hspace{0.75cm}+\Big( \langle\tilde{\varphi}^{\mu\alpha}_1, \varphi^\beta_0 \rangle \nabla_\mu
    +
    \langle \tilde{\varphi}^{\mu\alpha}_1, L^{-1} \varphi^{\nu\beta}_1 \rangle \nabla_\mu \nabla_\nu
    \Big) \delta X_\beta  = 0.\nonumber\\
  \end{eqnarray}
  Here, we have used the following simple relation
  \begin{eqnarray}
    \label{eq:5-009}
    \delta (f^{\mathrm{eq}}_p) = f^{\mathrm{eq}}_p  \varphi^\alpha_{0p}  \delta X_\alpha,
  \end{eqnarray}
  We note that
  all of the coefficients in Eq. (\ref{eq:5-008}) take a value of the
constant
solution
  $(T,\,\mu,\,u_\mu) = (T_0,\,\mu_0,\,u_{0\mu})$.
  Owing to the orthogonality  between the P${}_0$ and Q${}_0$ spaces,
   Eq. (\ref{eq:5-008}) is reduced to
  \begin{eqnarray}
    \label{eq:5-011}
    \eta^{\alpha\beta}_0\,\frac{\partial}{\partial\tau}\delta X_{\beta} +
    B^{\alpha\beta}\,\delta X_{\beta} = 0.
  \end{eqnarray}
Here $\eta^{\alpha\beta}_0$ is the metric tensor defined in (\ref{eq:1-043}) and 
$B^{\alpha\beta}$
is defined by
  \begin{eqnarray}
        \label{eq:5-013}
    B^{\alpha\beta} \equiv \langle \tilde{\varphi}^{\mu\alpha}_1,\varphi^\beta_0 \rangle
    \, \nabla_\mu + \eta^{\mu\alpha\nu\beta}_1 \, \nabla_\mu \, \nabla_\nu,
  \end{eqnarray}
with
   \begin{eqnarray}
    \label{eq:1-076}
    \eta^{\mu\alpha\nu\beta}_1 \equiv \langle \varphi^{\mu\alpha}_1, L^{-1} \varphi^{\nu\beta}_1  \rangle.
  \end{eqnarray}
 Both $\eta_0$ and $B$ are symmetric tensors.
  
  With the ansatz
$\delta X_{\alpha}(\sigma; \tau) =$
$\delta \tilde{X}_{\alpha}(k ; \Lambda) \mathrm{e}^{ik\cdot\sigma - \Lambda\tau}$,
  Eq. (\ref{eq:5-011}) leads to the following algebraic equation,
  \begin{eqnarray}
    \label{eq:5-015}
    ( \Lambda\,\eta^{\alpha\beta}_0 - \tilde{B}^{\alpha\beta} )
    \, \delta \tilde{X}_{\beta}
    = 0,
  \end{eqnarray}
with $\tilde{B}^{\alpha\beta} \equiv i\langle \tilde{\varphi}^{\mu\alpha}_1,\varphi^\beta_0\rangle 
 k_\mu -\eta^{\mu\alpha\nu\beta}_1 k_\mu k_\nu$.
Thus we have the eigenvalue equation as follows,
  \begin{eqnarray}
    \label{eq:5-017}
    \det ( \Lambda\,\eta_0 - \tilde{B} ) = 0,
  \end{eqnarray}
  which would give  the dispersion relation
 $\Lambda = \Lambda(k)$.
  The stability of the
generic constant
solution (\ref{eq:5-001})
  against a small perturbation
  is assured when
  the real part of $\Lambda(k)$ is nonnegative for any $k^\mu$,
which is shown to be the case as follows.
  
  Now, recall that the metric matrix $\eta_0$ is a real symmetric and positive-definite matrix, 
which implies that it has a Cholesky decomposition,
    \begin{eqnarray}
    \label{eq:5-020}
    \eta^{-1}_0 = {}^tU\,U,
  \end{eqnarray}
  where $U$ denotes a real matrix and ${}^tU$ a transposed matrix of $U$.
  Then Eq. (\ref{eq:5-017}) is converted to
  \begin{eqnarray}
    \label{eq:5-021}
    \det ( \Lambda \, I - U \, \tilde{B} \, {}^tU ) = 0,
  \end{eqnarray}
where $I$ denotes the unit matrix.
Eq. (\ref{eq:5-021}) tells us that 
  $\Lambda(k)$ is an eigen value of $U\, \tilde{B} \, {}^tU$.
  
  There is a  following theorem:
  The real part of the eigen value of a complex matrix $C$
  is nonnegative
  when the Hermite matrix $\mathrm{Re}(C) \equiv (C + C^\dagger)/2$
  is semi-positive definite.
  Applying this theorem to the present case,
  we find that
  the real part of $\Lambda(k)$ becomes nonnegative for any $k^\mu$
  when
  $\mathrm{Re}(U \, \tilde{B} \,{}^tU)$
  is a semi-positive definite matrix, which is shown to be the case, as follows;
    \begin{eqnarray}
    \label{eq:5-022}
    &&w_{\alpha} [ \mathrm{Re}(U  \tilde{B} {}^tU) ]^{\alpha\beta} w_{\beta}\nonumber\\
    &&\hspace{1cm}= w_{\alpha}  [ U \mathrm{Re}(\tilde{B})  {}^tU ]^{\alpha\beta} w_{\beta}\nonumber\\
    &&\hspace{1cm}= [w  U]_{\alpha} [\mathrm{Re}(\tilde{B})]^{\alpha\beta} [w  U]_{\beta}\nonumber\\
    &&\hspace{1cm}= - [w \, U]_{\alpha} \, \eta^{\mu\alpha\nu\beta}_1 \,
    k_\mu \, k_\nu \, [w \, U]_{\beta}\nonumber\\
    &&\hspace{1cm}= - \langle\, k_\mu\,[w \, U]_{\alpha} \, \varphi^{\mu\alpha}_1\,,\,
    L^{-1} \, k_\nu\,[w \, U]_{\beta} \, \varphi^{\nu\beta}_1
    \,\rangle\nonumber\\
    &&\hspace{1cm}= - \langle\, \psi \,,\,
    L^{-1} \, \psi \,\rangle \ge 0
    \,\,\,\mathrm{for}\,\,\,w_{\alpha} \neq 0,
  \end{eqnarray}
  with $\psi_p \equiv k_\mu\,[w \, U]_{\alpha} \, \varphi^{\mu\alpha}_{1p}$.
  This completes the proof that
  the
generic constant
solution in Eq. (\ref{eq:5-001}) is
  stable against a small perturbation.

\section{Second-order equations and moment method}

In  the first-order hydrodynamic equations, 
the zero modes of the linearized collision operator
 form the invariant manifold on which hydrodynamics is defined;
the would-be constant zero modes acquire the time-dependence on the manifold by the
RG equation. Our formalism can be extended so as to incorporate excited modes as additional
components of the invariant/attractive manifold \cite{next002}, and hence 
we can derive an extended thermodynamics or Israel-Stewart type equation
 with novel microscopic expressions 
of the relaxation times and lengths \cite{next002}. Furthermore,
our theory suggests a proper ansatz for the distribution function to be used in the
moment method \cite{next002}. 
For the shortage of space, we here give a sketch of some of our results
for the extended thermodynamics,
leaving the detailed derivation in a separate paper \cite{next002}.
We emphasize that our theory gives an explicit construction of 
the invariant manifold corresponding to thirteen moments, 
which has been long sought for \cite{meso,Gorban}.

\subsection{
A brief review of  Grad's thirteen-moment method 
and Grad-M\"{u}ller equation: non-relativistic case
}
\label{sec:002-2}
In Grad's thirteen-moment method \cite{grad,mic001},
the one-particle distribution function $f_{\Vec{v}}(t,\,\Vec{x})$ is 
represented as
\beq
f_{\Vec{v}}(t, \Vec{x}) = f^{\mathrm{eq}}_{\Vec{v}}(t, \Vec{x}) 
\big(1 + \Phi_{\Vec{v}}(t,\Vec{x})\big),
\eeq
where
$f^{\mathrm{eq}}_{\Vec{v}}$ denotes the Maxwell distribution function
and $\Phi_{\Vec{v}}$ the deviation from $f^{\mathrm{eq}}_{\Vec{v}}$
given by
\beq
\Phi_{\Vec{v}}(t,\Vec{x})&=&
 \hat{\pi}^{ij}_{\Vec{v}}(t, \Vec{x}) \pi^{ij}(t, \Vec{x})
  + \hat{J}^{i}_{\Vec{v}}(t, \Vec{x}) J^{i}(t, \Vec{x})\nonumber \\
 &\equiv& \Phi^{\mathrm{G}}_{\Vec{v}}(t, \Vec{x}),
\eeq
with
\begin{eqnarray}
\hat{\pi}^{ij}_{\Vec{v}}(t,\,\Vec{x}) &\equiv&
  m \,\Big(\delta v^i(t,\,\Vec{x}) \, \delta
  v^j(t,\,\Vec{x}) - \frac{1}{3} \, \delta^{ij} \, |\Vec{\delta v}(t,\,\Vec{x})|^2\Big),
\nonumber
  \\\\
\hat{J}^i_{\Vec{v}}(t,\,\Vec{x}) &\equiv&
  \Big(\frac{m}{2}\,
  |\Vec{\delta v}(t,\,\Vec{x})|^2 - \frac{5}{2}\,T(t,\,\Vec{x})\Big)\,\delta v^i(t,\,\Vec{x}).
\end{eqnarray}
Here,
$\Vec{\delta v}(t,\,\Vec{x}) \equiv \Vec{v} - \Vec{u}(t,\,\Vec{x})$ is 
the peculiar velocity.

Then the evolution equation of the thirteen coefficients
are determined by the equations all of which are derived from 
the Boltzmann equation
$(\partial/\partial t + \Vec{v}\cdot\Vec{\nabla})f_{\Vec{v}}(t,\,\Vec{x})
= C[f]_{\Vec{v}}(t,\,\Vec{x})$
with use of 
the linearized collision operator given by
\beq
L_{\Vec{v}\Vec{k}} \equiv (f^{\mathrm{eq}}_{\Vec{v}})^{-1}\,
   \frac{\partial}{\partial f_{\Vec{k}}}C[f]_{\Vec{v}}\Bigg|_{f=f^{\mathrm{eq}}} \,
   f^{\mathrm{eq}}_{\Vec{k}}.
\eeq

Thus, the Grad-M\"{u}ller equation
is obtained
as a closed system of the equations governing 
$T$, $n$, $u^i$, $\pi^{ij}$, and $J^i$ in terms of 
the transport coefficients and the relaxation times,
which  are given in terms of 
an inner product defined by
${\langle\, \psi\,,\, \chi\,\rangle}_{\mathrm{eq}} \equiv \sum_{\Vec{v}} \, f^{\mathrm{eq}}_{\Vec{v}}
   \, \psi_{\Vec{v}} \, \chi_{\Vec{v}}$.
For example, the shear viscosity and the relaxation time of the stress tensor 
$\hat{\pi}^{ij}$ are expressed as
\begin{eqnarray}
  \label{eq:moment-12}
  \eta^{\mathrm{G}} &=&
 - \frac{1}{10\,T}\,\frac{{\langle\, \hat{\pi}^{ij}\,,\,\hat{\pi}^{ij} \,\rangle}_{\mathrm{eq}}
  \, {\langle\, \hat{\pi}^{kl}\,,\,\hat{\pi}^{kl} \,\rangle}_{\mathrm{eq}}}{
  {\langle\, \hat{\pi}^{mn}\,,\,L \, \hat{\pi}^{mn} \,\rangle}_{\mathrm{eq}}},\\
  \tau^{\mathrm{G}}_{\pi} &=& -\frac{{\langle\, \hat{\pi}^{ij}\,,\,\hat{\pi}^{ij} \,\rangle}_{\mathrm{eq}}}
  {{\langle\, \hat{\pi}^{kl}\,,\,L\,\hat{\pi}^{kl} \,\rangle}_{\mathrm{eq}}}.
  \end{eqnarray}
It is well known that 
the formula (\ref{eq:moment-12}) is different from that
given in the Chapman-Enskog expansion method, and 
there are many attempts both in non-relativistic  \cite{Gorban}
and relativistic cases \cite{mic004,Denicol:2010xn}
 to modify and/or extend the Grad moment method 
so that the transport coefficients thus obtained become consistent with those
obtained by Chapman-Enskog method.

\subsection{Relativistic mesoscopic dynamics from the RG method}

In this subsection, we make a brief report on our
attempt \cite{next002} to extend the RG method so as to obtain the so called 
mesoscopic dynamics \cite{meso} in the relativistic case.

First we show the results in such  a way that a comparison with the 
Grad moment method  is apparent.
If we express the distribution function by
$f_p(x) = f^{\mathrm{eq}}_p(x)\,\big(1 + \Phi_p(x)\big)$,
our RG method gives the following expression 
of $\Phi_p(x)$,
\begin{eqnarray}
  \Phi_p = -\frac{1}{T}\,\sum_q
  \, L^{-1}_{pq}\,
  \Big(
    \tilde{\Pi}_q \, \frac{\Pi}{\zeta}
  + \tilde{J}^\mu_q \, \frac{J_\mu}{\lambda}
  + \tilde{\pi}^{\mu\nu}_q \, \frac{\pi_{\mu\nu}}{2\,\eta}
  \Big),\nonumber\\
\end{eqnarray}
where $\tilde{\Pi}_p$, $\tilde{J}^\mu_p$,
and $\tilde{\pi}^{\mu\nu}_p$ are the microscopic thermal forces.
This new form is different from any proposals 
in the literature \cite{mic004,Denicol:2010xn}.

The resultant energy-momentum tensor and particle current are found \cite{next002} to have 
the following forms,
respectively,
\beq
T^{\mu\nu}_{\mathrm{2nd}} &=& e \,u^\mu \,u^\nu - (p + \Pi) \,\Delta^{\mu\nu} + \pi^{\mu\nu}, \\
N^\mu_{\mathrm{2nd}} &=& n\,u^\mu + J^\mu.
\eeq
The relaxation equations derived in our RG method read
\begin{eqnarray}
  \Pi &=& -\zeta\,\nabla\cdot u
- \tau_\Pi \,  D\Pi\nonumber \\
 &&{} +
\textrm{other terms involving relaxation lengths},\\
  J^\mu &=&
\lambda\,\frac{1}{\hat{h}^2} \, \nabla^\mu \frac{\mu}{T}
  - \tau_J \, \Delta^{\mu a} \, DJ_a \nonumber \\
 &&{}+
\textrm{other terms involving relaxation lengths},\\
  \pi^{\mu\nu} &=&
2\,\eta\,\Delta^{\mu\nu\rho\sigma} \, \nabla_\rho u_\sigma
- 
\tau_\pi \, \Delta^{\mu\nu ab} \, D\pi_{ab} 
 \nonumber\\
  &&{}+ \Big(\kappa^{(0)}_{\pi\pi}\,\Delta^{\mu\nu\rho\sigma}\,\nabla\cdot u
  + \kappa^{(1)}_{\pi\pi} \, \Delta^{\mu\nu ac}\,
  \Delta_c^{\,\,\,b\rho\sigma} \, \Delta_{abde}\,\nabla^d u^e \nonumber \\
 &&{}+ \kappa^{(2)}_{\pi\pi}\,\Delta^{\mu\nu ac}\,
  \Delta_c^{\,\,\,b\rho\sigma} \, \omega_{ab}\Big) \, \pi_{\rho\sigma}\nonumber \\
 &&{}+\textrm{other terms involving relaxation lengths},
\end{eqnarray}
where
$\omega^{\mu\nu} \equiv \frac{1}{2} \, (\nabla^\mu u^\nu - \nabla^\nu u^\mu)$
is the vorticity.

Our RG method \cite{next002} gives microscopic expressions for the relaxation times
 $\tau_\Pi$, $\tau_J$, and $\tau_\pi$ as follows
\begin{eqnarray}
  \tau_\Pi &\equiv&
  - \frac{
    \langle \, \tilde{\Pi}\,,\,L^{-2}\,\tilde{\Pi} \, \rangle
  }{
    \langle \, \tilde{\Pi}\,,\,L^{-1}\,\tilde{\Pi} \, \rangle
  }
  =\frac{
    \int_0^\infty\!\!\mathrm{d}s\,\,\,s\,R_\zeta(s)
  }{
    \int_0^\infty\!\!\mathrm{d}s\,\,\,R_\zeta(s)
  }  ,
\eeq
and so on for 
$\tau_J$ and $\tau_{\pi}$ with obvious modifications.
We note that
our novel formulae for the relaxation times 
are all nicely represented in terms of the relaxation functions $R_\zeta(s)$, $R_\lambda(s)$, and
$R_\eta(s)$ so that they have a natural  physical meaning of
the relaxation time as the  correlated time of the respective relaxation function
 in contrast to other approaches \cite{mic004,grad,Denicol:2010xn}.
We also mention that bulk and shear viscosities and the heat conductivity derived in our method do 
coincide with those in the Chapman-Enskog method as shown before.

\section{Summary and concluding remarks}

We have reported our attempts to derive first-order and second-order
relativistic hydrodynamic equations from  relativistic Boltzmann equation which 
has a manifest Lorentz invariance and does not show any pathological 
behavior such as the instability and acausality seen in existing hydrodynamic equations.
We have given an argument, on a physical ground on the nature of the origin of the 
dissipation and the form of the local equilibrium distribution function, 
that the energy frame is uniquely chosen as the one
 in which the relativistic hydrodynamic equation
for a viscous fluid is defined.
We have given the novel extended thermodynamics both in non-relativistic and
relativistic cases through the explicit construction of attractive manifold containing
the relaxation process from Boltzmann equation. 

  It is worth emphasizing that  that  all the equations derived in this work
  are consistent with the underlying kinetic equation, \textit{i.e.},
relativistic Boltzmann equation.
This is one of the advantage in our theory because
our theory explicitly gives the solution (distribution function) 
of  the Boltzmann equation, which is expressed  with the hydrodynamic variable and relaxation times and lengths,  and thereby
makes a systematic description of the time-evolution of the system from 
 hydrodynamic to kinetic regime.  Such an  overall analysis should be desirable for 
 that of the matter created at RHIC, LHC and other 
systems   where the proper dynamics would change
  from the hydrodynamic to the kinetic ones or vice versa
 \cite{Hirano:2012kj,connect-1,connect-2,Hirano:2005wx,dis005,nonaka06}.
Furthermore,
it would be interesting to evaluate the relaxation times as well as the transport coefficients
of the created matter
with the use of
the microscopic representations obtained in this work.

  Finally,
  we note that the renormalization-group method \cite{rgm001,env001,env006,env007}
 itself has
  a universal nature
  and can be applied to derive a slow dynamics
  from kinetic equations other than the simple Boltzmann equation, say, 
Kadanoff-Baym equation \cite{kadanoff}.

\section*{Acknowledgments}

T.K. thanks the editors to invite him to contribute to this special issue in 
Euro Physics A.
  We are grateful to K. Ohnishi for his collaboration in the early stage of this work.
  T.K. was partially supported by a
  Grant-in-Aid for Scientific Research from the Ministry of Education,
  Culture, Sports, Science and Technology (MEXT) of Japan
  (Nos. 20540265 and 23340067),
  by the Yukawa International Program for Quark-Hadron Sciences, and by a
  Grant-in-Aid for the global COE program
  ``The Next Generation of Physics, Spun from Universality and Emergence'' from MEXT.
%

\end{document}